\documentclass[twoside]{report}
\usepackage{iwsm}
\usepackage{graphicx}
\usepackage{amsmath, amssymb}
\usepackage{booktabs}

\begin{document}

\title{Mean survival time by ordered fractions of population with censored data}
\titlerunning{Mean survival by ordered fractions with censoring}

\author{Celia Garc\'{i}a-Pareja\inst{1}, Matteo Bottai\inst{1}}
\authorrunning{Garc\'{i}a-Pareja and Bottai}

\institute{Unit of Biostatistics, Institute of Environmental Medicine, Karolinska Institutet, Stockholm, Sweden}

\email{celia.garcia.pareja@ki.se}

\abstract{We propose a novel approach for estimating mean survival time in the presence of censored data, in which we divide the 
population under study into survival-ordered fractions defined by a set of proportions, 
and compute the mean survival time for each fraction separately. 
Our approach provides a detailed picture of the distribution of the time variable while preserving 
the appealing interpretation of the mean.
Our measure proves to be of great use in applications, particularly those where we are able to 
detect differences in mean survival across groups for certain fractions of the population that would 
have been overlooked using other available methods.}


\maketitle

\section{Introduction}
Survival data analysis methods are the cornerstone of a wide range of statistical applications. While mean survival time is 
of utmost relevance, e.g., 
in health economics (Paltiel et al., 2009) 
or oncology studies (Zhao et al., 2001), its estimation might be hindered in the presence of censoring, 
where the time variable is only observed until a certain quantile. In practice, censoring is almost 
always present, calling for specialised  estimation techniques. Several approaches have been considered 
to overcome this problem, the most 
used amongst them, the restricted mean, computes mean survival time up to a specific cut-off time 
point (Irwin, 1949).
Estimation of the restricted mean, however, might be heavily affected by the presence of censored observations, which will 
result in a loss of estimation accuracy. 
Moreover, clinical relevance and interpretation of restricted mean estimates remains unclear.

We present a novel mean survival measure based on observed quantiles that divides the population 
in ordered fractions in which the mean survival can be estimated separately. Interpretation of 
the estimates is straightforward, as they represent mean survival 
times for the specified fractions of the population. Similarly to the restricted mean, 
we estimate mean survival up to a specific cut-off point, 
that we set to the largest observed $p$-th fraction of population to experience the event of interest. 
Our approach exploits that the distribution of observed and censored events imposes differences 
in the estimation accuracy of specific quantiles, i.e., those that are close to observed events can be more precisely estimated than 
those located after the occurrence of censored events. 
Therefore, estimates for certain fractions can be really precise, 
which allows quantifying significant mean survival differences across groups, even in scenarios where state-of-the-art 
methods are unable to 
detect them.

\section{Mean survival by ordered fractions}
Let $T$ be a non-negative random variable with $\text{E}[T]<\infty$ and let $S(\cdot)$ and $Q(\cdot)$ denote its survival 
and quantile functions, respectively. An expression for the expectation of $T$ in terms of $Q(\cdot)$ is

\begin{equation}
\mu=\text{E}[T]=\int_0^{\infty} S(t) \mathrm{d}t = \int_0^{1} Q(p) \mathrm{d}p. \label{garciapareja:mu_quantile}
\end{equation}

Given a grid of proportions $\{\lambda_0,\lambda_1,\ldots,\lambda_K\}$ with $\lambda_{k-1}< \lambda_k$ for all 
$k\in\{1,\ldots,K\}$, we can divide $\mu$ into separate components as follows

\begin{equation}
\mu=\sum_{k=1}^K \mu_k, \text{ where }  \mu_k= \int_{\lambda_{k-1}}^{\lambda_k} Q(p) \mathrm{d}p, \lambda_0=0 \text{ and } \lambda_K=1. \label{garciapareja:mu_decomp}
\end{equation}

If we now weight each $\mu_k$ by its corresponding inverse proportion, we obtain
$$
\overline{\mu}_k= \frac{\mu_k}{\lambda_k-\lambda_{k-1}},
$$
where $\overline{\mu}_k$ is the mean survival time for a specific fraction of population delimited by $(\lambda_{k-1},\lambda_{k})$. 
For example, 
if we consider $(\lambda_0,\lambda_1)=(0,0.5)$, $\overline{\mu}_1$ quantifies mean survival time for the first half of the population to experience 
the event of interest.

In the presence of a censoring variable $C$, when $Y=\min(T,C)$ is observed, the decomposition shown 
in (\ref{garciapareja:mu_decomp}) 
is of utmost convenience because $\lambda_K$ can be set to the largest proportion of observed events, 
that is, the one corresponding to the last observed quantile.
Note that when $\lambda_K<1$, the mean survival time for the $\lambda_K$-th fraction of the population observed to experience the event, 
does not correspond to the restricted mean computed up to the last observed quantile $y^{\star}=Q(\lambda_K)$.
Indeed, while
$$
\overline{\mu}_K=\frac{1}{\lambda_K}\int_{0}^{\lambda_K} Q(p) \mathrm{d}p
$$
can be easily interpreted in terms of the population under study, the corresponding
$$
\mu^{\star}=\int_{0}^{y^{\star}} S(y) \mathrm{d}y,
$$ 
does not prove as informative.

\section{Estimation and simulation results}

In the presence of censoring, estimation of $\mu_k$ is possible via the Kaplan-Meier estimator of the underlying 
survival function, $\widehat{S}(\cdot)$. Given $\widehat{S}(\cdot)$ and the grid of proportions 
$\{\gamma_0,\gamma_1,\ldots,\gamma_K\}=\{1-\lambda_0,1-\lambda_1,\ldots,1-\lambda_K\}$, an 
estimator for $\mu_k$ follows easily from equations (\ref{garciapareja:mu_quantile}) and (\ref{garciapareja:mu_decomp}), with
$$
\begin{array}{rl}
 \widehat{\mu_k}
 &=\displaystyle \sum_{j=1}^{J_k} y_{j} [ \min \{ \widehat{S}( y_{j-1} ) , \gamma_{k-1}\}- \max \{ \widehat{S}( y_{j} ), \gamma_k \}]\\
&=\displaystyle\sum_{j_k=1}^{{J_k}} \widehat{Q}(p_{j}) (p_{j}-p_{j-1}),
\end{array}
$$
where $y_{j}$ denote observed event times such that $\widehat{S}(y_{j})\in [\gamma_{k}, \gamma_{k-1}]$ for all $j\in\{1,\ldots, J_k\}$, 
and $\widehat{S}(y_0)\geq \gamma_{k-1}$ and 
$\widehat{S}(y_{J_k})\leq \gamma_{k}$. In this case, we obtain a step-wise constant estimator of the quantile function $\widehat{Q}(\cdot)$, in which 
observed times $y_{j}$ play the role of estimated quantiles $\widehat{Q}(p_{j})$ of order $p_{j}=\widehat{S}( y_{j} )$.

We tested the performance of $\widehat{\mu_k}$ in different scenarios, all yielding analogous conclusions. 
In Table~\ref{garciapareja:simulation} we present results for a simulation study of $5,000$ data sets with $200$ samples each,
generated from a time variable following 
a log-logistic distribution with scale $\alpha=1$ and shape $\beta=2$. The censoring variables were sampled independently from a 
uniform distribution in $(0,7/3)$, 
yielding an average censoring rate of $50\%$. Estimated average upper and lower bounds for $\widehat{\mu_k}$ where computed integrating 
over equal precision confidence bands for the Kaplan-Meier estimator (Nair, 1984).
We observed that our estimates' precision decreased with increasing $k$ (that is, the bands widened with increasing $k$), which was expected, 
as the proportion of censored observations also increased with $k$ and fewer events were observed. In this sense, 
one might say that some $\mu_k$ can be \textit{more precisely} estimated than others, which proves highly useful in application seetings. 

\begin{table}[!ht]\centering
\caption{\label{garciapareja:simulation} Results for $5,000$ samples of $200$ observations from a log-logistic model with scale $\alpha=1$ and shape $\beta=2$ with censoring variable uniform in ($0$,$7/3$), 
corresponding to an average censoring rate of $50\%$. True ($\mu_k$) and average estimated ($\widehat{\mu_k}$) values, 
with average lower ($\widehat{\mu^L_k}$) and 
upper ($\widehat{\mu^U_k}$) bounds for $5$ fractions of 
population, $\%$ of simulations ($\text{nsim}_k$) in which $\widehat{\mu_k}$ could be computed and average number of observed events 
($\text{d}_k$) are reported. The average bound marked with $^{\star}$ had finite values in $75\%$ of the simulations.}
\medskip
\begin{tabular}{ccccccc}
\toprule[0.09 em]
$k$&$\lambda_k$& $\mu_k$ & $\widehat{\mu_k}$ & $\widehat{\mu^L_k}-\widehat{\mu^U_k}$ & $\text{nsim}_k$ & $\text{d}_k$ \\
 \midrule
 $1$&$0.20$ & $0.064$  &  $0.064$ & $0.044- 0.086$ & $100\% $ & $34$   \\
 $2$&$0.40$ & $0.131$ &   $0.132 $&  $0.101- 0.175 $ & $100\% $ & $29$\\
 $3$&$0.60$ & $0.201$ &  $0.202  $&  $0.156-0.264^{\star} $ & $100\% $ & $23$\\
 $4$&$0.80$& $0.311$ &   $0.304  $&  $0.226-\infty $ & $70.7\% $& $14$ \\
 $5$&$0.95$& $0.420$ & $0.307 $ &$0.239-\infty$ & $5.80\% $ & $4$\\
\bottomrule[0.09 em]
\end{tabular}
\end{table}

\section{Application example: Survival after bone marrow transplant in lymphoma patients}
\begin{figure}[!ht]\centering
\includegraphics{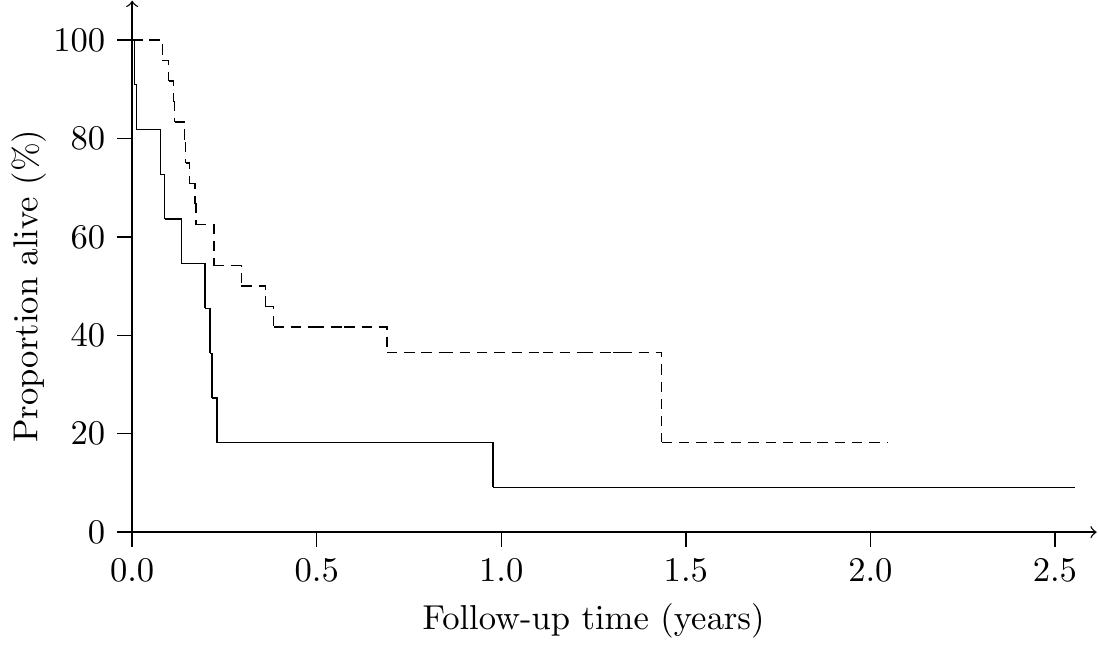}
\caption{\label{garciapareja:survivalcurve} Estimated Kaplan-Meier survival curves after bone marrow transplant 
for lymphoma patients that received allogenic 
(solid line) or autologus 
(dashed line) transplant.}
\end{figure}

We analysed data on 35 patients with lymphoma that received either an allogenic or an autologous bone marrow transplant, 
that is, they received marrow from either a compatible donor or their own after chemotherapy treatment and cleansing 
(Avalos et al., 1993).
The aim of the study was to find differences between lymphoma-free survival after having received either type of transplant. 
After 2.5 years of follow-up, 26 patients had died or relapsed and the 
censoring rate was $25.7\%$. The estimated survival curves for both treatments are shown in Figure~\ref{garciapareja:survivalcurve}.

While restricted mean survival estimates did not detect any significant difference in mean 
survival between the allogenic and autologus transplant groups (restricted mean difference of $146.5$ days, with $95\%$ confidence interval $(-29.71, 322.7)$)), 
our approach showed that among earlier failures that difference was actually 
significant. In particular, considering the weakest $10\%$ of the patients, that is, the first $10\%$ to die or relapse after receiving 
the transplant, mean survival difference was estimated at $32.15$ days ($95\%$ CI $(13.98-50.31)$) favouring those who received the autologus 
transplant.
In Table~\ref{garciapareja:dataresults} we show the results of mean survival time differences after receiving a bone marrow transplant  
by deciles of population up to the $80$th percentile (last fraction commonly observed in both groups). 
\begin{table}[!ht]\centering
\caption{\label{garciapareja:dataresults} Estimates for mean survival differences between 
allogenic ($\widehat{\overline{\mu^0}_k}$) and autologus ($\widehat{\overline{\mu^1}_k}$) bone marrow transplants 
with bootstrapped $95\%$ confidence intervals by ordered deciles of population.}
\medskip
\begin{tabular}{cccc}
\toprule[0.09 em]
$k$&$\lambda_k$& $\widehat{\overline{\mu^1}_k}-\widehat{\overline{\mu^0}_k}$ & $95\%$ CI\\
 \midrule
 $1$&$0.1$ & $32.15$  & $13.98-50.31$ \\
 $2$&$0.2$ & $36.72$ & $2.843- 70.60$\\
 $3$&$0.3$ & $26.23$ & $-19.94- 72.43$\\
 $4$&$0.4$& $28.98$ & $-40.51- 98.48$\\
 $5$&$0.5$&  $32.80$ & $-124.5- 190.1$\\
 $6$&$0.6$&  $80.60$ & $-1283- 1444$\\
 $7$&$0.7$&  $349.4$ & $-446.4- 1145$\\
 $8$&$0.8$&  $441.4$& $-130.3- 1013$\\
\bottomrule[0.09 em]
\end{tabular}
\end{table}
Our estimates could detect an improvement on lymphoma-free survival for the autologus transplant group amongst at least the weakest $20\%$ of patients, providing a useful 
guide for effective decision-making.

\section{Final remarks}
Our approach for quantifying mean survival time takes advantage of the information contained in the data and deals with the censoring hurdle. 
Our proposed measures are easily interpretable, providing a useful alternative to the restricted mean, which poses interpretation difficulties.
By dividing the study population into ordered fractions, the proposed method provides a detailed picture of the underlying probability distribution and can 
detect mean survival differences across groups that are often undetected by other methods. 
Results from a simulation study show good performance of our proposed 
estimation strategy even in the presence of censoring, and support the idea that mean survival 
can be more accurately estimated in some fractions of the population. 
In the analysis of survival data after bone marrow transplant, our method detected differences in mean survival between given transplants 
for certain fractions of population, while those differences were overlooked when using restricted mean estimates instead.

\acknowledgments{This work was partially supported by the KID funding doctoral grant from Karolinska Institutet}

\references
\begin{description}
\item[Avalos, B.R., Klein, J.L., Kapoor, N., Tutschka, P.J., et al.] (1993).
     Preparation for Marrow Transplantation in Hodgkin's and non-Hodgkin's Lymphoma Using Bu/CY.
     {\it Bone Marrow Transplantation}, {\bf 12},
      133\,--\,138.
\item[Irwin, J. O.] (1949).
     The Standard Error of an Estimate of Expectation of Life, with Special Reference to Expectation of 
     Tumourless Life in Experiments with Mice.
     {\it Journal of Hygiene}, {\bf 47},
      188\,--\,189.
\item[Nair, V.N.] (1984).
     Confidence Bands for Survival Functions with Censored Data: A Comparative Study
     {\it Technometrics}, {\bf 26},
      265\,--\,275.
\item[Paltiel A.D., Freedberg K.A., Scott C.A., Schackman B.R., Losina E. et al.] (2009).
     HIV Preexposure Prophylaxis in the United States: Impact on Lifetime Infection Risk, Clinical Outcomes, and Cost-Effectiveness.
     {\it Clinical Infectious Diseases}, {\bf 48},
      806\,--\,815.
\item[Zhao Y., Zeng D., Socinski M.A., Kosorok M.R.] (2011).
     Reinforcement Learning Strategies for Clinical Trials in non-Small Cell Lung Cancer.
     {\it Biometrics}, {\bf 67},
      1422\,--\,1433.
\end{description}

\end{document}